# The Impact of Mobility Model in the Optimal Placement of Sensor Nodes in Wireless Body Sensor Network

B. O. Sadiq, A. E Adedokun, Z.M Abubakar
Department of Electrical and Computer Engineering
Ahmadu Bello University,
Zaria Nigeria
bosadiq@abu.edu.ng, wale@abu.edu.ng, xeenabu@yahoo.com

***Abstract*—Power and energy consumption is a fundamental issue in Body Sensor Networks (BSNs) since nodes must operate properly and autonomously for a certain period of time without battery replacement or change. This is due to the fact that the sensors in BSNs are either implanted in the body or are in very near position to the body. Thus, the duration of replacing the batteries should be of utmost importance. Most of the existing researches suggested the development of a more improved battery cells or developing an energy aware routing protocol to tackle the energy consumption in WBSN. But this is not the case as most energy consumption in WBSN occur as a result of mobility in routing and sensor node placement. Therefore, improving the battery cells might not solve the energy consumption in WBSN.**

## I. Introduction

Body Area Network (BAN) can be defined as a system of devices in close proximity to a person's body, with a view to coordinating and cooperating for the benefit of the user. The BAN is still a developing idea that arose as a result of byproduct of current biomedical engineering and sensor network technology [1]. The wireless BAN, also referred to as Wireless Body Area Network (WBAN) or a Wireless Body Sensor Network (BSN) was coined out by Professor Guang-Zhong Yang in the year 2006, as a network of wearable computing devices.

Wireless Body Sensor Network (WBSN) can be defined as a self-directed, self-sufficient and self-governing i.e. autonomous system which is used to monitor the daily life activities of a person i.e., it can be effective in the monitoring and analysis of gesture detection, emotional recognition amongst others [2]. Thus, the WBSN is said to be a communication network between the human and the computer through wearable devices, where the interactions can be established using either an unguided sensor network or ad hoc network [3]. These wearable devices are small and intelligent, usually rooted in the body. They are also capable of establishing a wireless communication link [4]. The wireless devices can be categorized as either a sensor employed with a view to quantifying certain parameters of the human body, either externally or internally e.g. measurement of heartbeat, body temperature or recording a continued electrocardiogram (ECG) or an actuator used for precise actions according to the data they receive from the sensors or through interactions with the user [5].

The unguided nodes that are very resource restricted and poses hard to meet requirements in the aspect of computational and storage resources, battery duration, wearability amongst others are the factors that affect the medium of wireless communication between devices in WBSN [6]. Therefore, a WBSN architecture needs to consider large number of factors which might lead to improve network lifetime, lesser communication error and secure & safe communication medium.

The WBSN architecture is said to be dependent on the Energy, QoS & Reliability, Co-Existence, security & privacy, optimal placement of sensor nodes and data validation amongst others [7].

According to the early researchers in WBSN, its architecture, as in Figure 1, was faced with varieties of issues such as [8]:

i. Physical layer issues: which deal with the band selection, fault tolerance and interference
ii. MAC layer issues: which deal with reliability, dynamic channel assignment and scheduling

| Challenges | Differences between the WSN and WBSN | |
|---|---|---|
| | ***Wireless Body Area Network*** | ***Wireless Sensor Network*** |
| Node Number | Fewer, limited in space | Several redundant nodes for extensive range coverage |
| Scale | Human body (centimeters / meters) | Monitored environment (meters / kilometers) |
| Result accuracy | Done using node exactness and robustness | Through node redundancy |
| Node Tasks | Node performs multiple tasks | Node performs a dedicated task |
| Node Size | Large is not preferred. i.e. better small | Small is preferred, nonetheless significant |
| Node Replacement | Replacement of implanted nodes difficult | Performed easily, nodes even disposable |
| Node Lifetime | Some years / months, smaller battery capacity | Some years / months |
| Power Supply | It will be a herculean task to replace in a rooted setting | Accessible and likely to be replaced more easily and frequently |
| Security Level | Very high | Lower |
| Effect of Information Forfeiture | More significant, may require additional measures to ensure QoS and real-time data delivery | Probable to be compensated by redundant nodes |

iii. Routing issues: which involve the varying data needs, resource constraints, mobility amongst others.
iv. Energy consumption and network life time improvement.

However, the major issue considered by many researchers is the energy consumption with a view to improving the network life time. Most of the existing researches suggested the development of a more improved battery cells or developing an energy aware routing protocol to tackle the energy consumption in WBSN. But this is not the case as most energy consumption in WBSN occur as a result of mobility in routing. Therefore, improving the battery cells might not solve the energy consumption in WBSN.

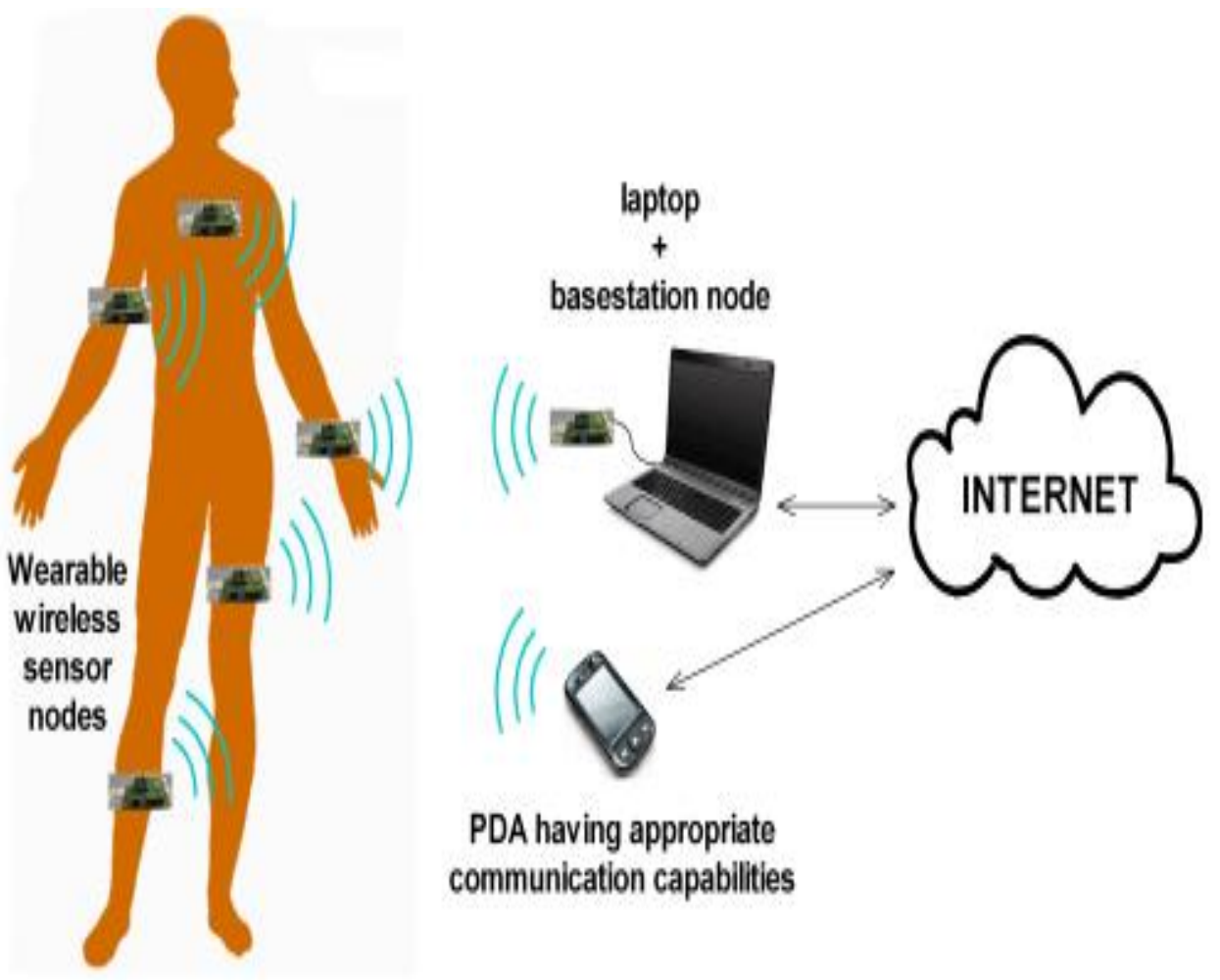

Figure 1: A Typical WBSN Architecture

The optimal placement of sensor nodes in Wireless Sensor Network (WSN) is different to that of the Wireless Body Sensor network (WBSN) due to the inherent characteristics they possess. The difference between them are summarized in Table 1

Table 1: Difference between WBSN and WSN [6] [9]

There exist different types of wireless body sensor network presented in literature such as [10]:

i. Managed WBSN: This type of WBSN usually involves a third party which is either the health personnel or the medical center that will decide on the information collected from one or more sensor nodes. The third party as full control over the data this is collected.
ii. Autonomous WBSN: The Autonomous WBSN is similar to the Managed WBSN but with the presence of actuators. The actuators in collaboration with the sensor nodes causes action on the human body. This action is as a result of the data collected from the sensor nodes by direct interaction with the human body without involving an intermediary. The actuators are capable of taking decisions in real time without delay.
iii. Intelligent WBSN: Intelligent WBSNs are combination of both the Managed WBSN and the Autonomous WBSN. Decisions can be taken by both the actuators and the third party depending on the complexity of the information [10].

Therefore, in WBSN architecture, due to the fact that the different nodes perform varieties of tasks, node cooperation is required. There are advantages of using node cooperation. Some of which are:

i. Speed of acquiring and disseminating information: Different tasks can be completed faster with more than one node, since tasks will be split and the information obtained will be shared.
ii. Enhancing communication since nodes are fewer and limited in space.
iii. Increasing node life time since battery capacity are smaller.
iv. Minimizing the impact of data loss to the barest minimum since there are no redundant nodes to compensate for data loss as compared to the WSN architecture.

## II. LITERATURE REVIEW

Reducing energy consumption with a view to increasing network life time has been presented by a number of researchers such as the work of [8], [11] and [12]. These researchers presented the issue of energy consumption as the only WBSN parameter that requires optimization. Energy consumption is one of the underlaying blocks since Body Sensor Networks (BSNs) must operate appropriately and autonomously for a certain period of time without battery replacement or change. Most of the existing researches suggested the development of a more improved battery cells or developing an energy aware routing protocol to tackle the energy consumption in WBSN. But this is not the case as most energy consumption in WBSN occur as a result of mobility in routing. Therefore, improving the battery cells might not solve the energy consumption in WBSN.

Early researches such as the work of [1] presented the concept and challenges in the WBSN. The challenges faced in BSNs is said to be similar to the general wireless sensor networks (WSN). However, an inquiry about the smaller and fewer nature of nodes relative to the conventional WSN triggered new research and design questions. Smaller nodes will use smaller batteries which in turn create some strict tradeoffs between the energy consumed by processing, communication and storage, resources and the fidelity, latency and throughput required by applications. Nonetheless, the key function of the WBSN is to effectively transmit and transform sensed phenomena into valuable information. This is expected to be done to meet other system requirements, such as storage, communication resources amongst others. This lead to the limitation of the work of [8], [11] and [12] which only considered improving battery cells and developing energy aware protocol in order to improve energy efficiency. In view of this, many researchers have tried to improve the wearability of sensor nodes by studying in-depth problems such as fault diagnosis, energy control and number of sensor nodes. However, most if not all the works encouraged the development of the WBSN in the direction of improving the battery cells or developing an energy aware routing protocol in order to improve energy efficiency.

Hence, the objective of this paper is to present the need to consider the impact of mobility model in the optimal placement of sensor nodes in the WBSN. This is because, the human body usually maintains three postures such as the standing, walking and running. But, has the human body moves either by walking or running, distance between the sensor nodes and sink increases or decreases. Thus, affecting the communication between nodes. This causes delay which in turns lead to energy consumption.

## III. RESEARCH PROBLEMS

Energy consumption in WBSNs happens during sensing, processing and communication as the early researchers have presented. Optimizing number and placement of nodes will improve communication, data processing and sensing whilst reducing the energy consumption and increasing the network life time. In WBSNs, nodes are placed in the human body to monitor heartbeat, pressure, temperature amongst others. There is a destination node called the sink, located in the chest of the human body were nodes sends their sensed data. The numbers and size of sensors in the human body is smaller compared to the general WSN network and the distance between the source nodes and the sink is constant in fixed position. But, has the human body moves either by walking or running, distance between the sensor nodes and sink increases or decreases.

In this section, we list a few still open research issues regarding the WBSN.

### *A.* Imapct of mobility models in sensor node placement

Optimizing number and placement of nodes will improve communication, data processing and sensing whilst reducing the energy consumption and increasing the network life time. In an optimal design of WBSN, there is a need to cope with several conflicting metrics while satisfying the given constraints. Optimizing a design to meet design requirements by maximizing or minimizing the metrics must take constraints such as mobility into account with a view to achieving an optimal result.

### *B.* Routing and protocols

Efficient and reliable communication protocol that considers the mobility is a critical issue to consider. A communication protocol that supports frequent retransmission of packets will shortens the life span of a node. There is need for seamless communication between nodes by implementing an on-demand protocol that will respond only when there is an information to be shared between nodes or from source to sink.

Therefore, in order to improve the node sensor placement problem and routing issues in WBSN, there is a need to

i. Develop an improved mobility model for node placement in the human body. The mobility model should consider three postures such as the standing, walking and running. A scenario in which all sensor nodes stream data towards sink nodes should be used.
ii. Develop an improved decision tree for routing protocol based on the three posture of the human body which are the standing, walking and running.

## IV. CONCLUSION

Energy consumption is one of the underlaying blocks since Body Sensor Networks (BSNs) must operate appropriately and autonomously for a certain period of time without battery replacement or change. This is due to the fact that the sensors in BSNs are either implanted in the body or are in very close proximity to the body thus, replacing the batteries should not be of utmost importance. Therefore, there is a need to consider mobility model for the optimal placement of sensor nodes and routing protocols in a wireless body sensor network with a view to improving the communication between nodes, maximizing throughput and reducing delay whilst guaranteeing energy efficiency and network life time.